%% file: apssamp.tex
\documentclass[
reprint,
superscriptaddress,
 amsmath,amssymb,
 aps,
]{revtex4-2}

\usepackage{graphicx}% Include figure files
\usepackage{subcaption}%subcation
\usepackage{caption}
\usepackage{xr}
\usepackage{xcolor}% for coloring parts of the template during development
\usepackage{dcolumn}% Align table columns on decimal point
\usepackage{bm}% bold math
\usepackage[version=4]{mhchem}
\usepackage{verbatim}
\usepackage[UKenglish]{babel} %BE hyphenation

\usepackage{booktabs}

\usepackage{lineno}  %for linenumbers
\usepackage{placeins}
\usepackage{siunitx}
\usepackage{ragged2e}

\usepackage{todonotes}
\usepackage{mhchem}

\usepackage{mathtools}

\DeclarePairedDelimiter\ket{\lvert}{\rangle}
\DeclarePairedDelimiterX\braket[2]{\langle}{\rangle}{#1 \delimsize\vert #2}

\begin{document}

\preprint{APS/123-QED}

\title{A diode nanocavity for fast, efficient and tunable emission of highly entangled photon pairs and Fourier-transform-limited single photons}
%(based on an electrically controlled QD in a CBR)
% Force line breaks with \\

%\begin{comment}
\author{Ievgen Brytavskyi
\textsuperscript{\S}
}
\thanks{These authors contributed equally to this work.}
%\thanks{Corresponding author: ievgen.btytavskyi@jku.at}
%\email{ievgen.brytavskyi@jku.at}

\author{Thomas Oberleitner}
\thanks{These authors contributed equally to this work.}
\author{Christian Weidinger}
\thanks{These authors contributed equally to this work.}
\author{Maximilian Aigner}
\author{Gabriel Undeutsch}
\author{Tobias Steindl}
\author{Johannes Reindl}
\author{Ailton Garcia Jr.}
\author{Melina Peter}

\author{Christian Schimpf}
\thanks{Current address: University of Cambridge, Cavendish Laboratory, J J Thomson Ave, Cambridge CB3 0HE, United Kingdom}

\author{Santanu Manna}
\thanks{Current address:
 Department of Electrical Engineering, Indian Institute of Technology Delhi, India
}%
\affiliation{%
 Institute of Semiconductor and Solid State Physics, Johannes Kepler University Linz, Altenberger Straße 69, 4040 Linz, Austria% \textbackslash\textbackslash
}%

\author{Michele B. Rota}
\affiliation{%
 Dipartimento di Fisica, Sapienza Università di Roma, Piazzale Aldo Moro 5, 00185 Roma, Italy% \textbackslash\textbackslash
}%
\author{Quirin Buchinger}

\author{Sven Höfling}
\affiliation{Julius-Maximilians-Universität Würzburg, Physikalisches Institut and Würzburg-Dresden Cluster of Excellence ctd.qmat, Am Hubland, 97074 Würzburg, Deutschland}

\author{Tobias Huber-Loyola}
\affiliation{Julius-Maximilians-Universität Würzburg, Physikalisches Institut and Würzburg-Dresden Cluster of Excellence ctd.qmat, Am Hubland, 97074 Würzburg, Deutschland}%
\affiliation{Institute of Photonics and Quantum Electronics (IPQ), Karlsruhe Institute of Technology (KIT), Engesserstr. 5, 76131 Karlsruhe, Germany}

\author{Rinaldo Trotta}
\affiliation{%
 Dipartimento di Fisica, Sapienza Università di Roma, Piazzale Aldo Moro 5, 00185 Roma, Italy% \textbackslash\textbackslash
}%

\author{Tobias M. Krieger}
\affiliation{%
 Institute of Semiconductor and Solid State Physics, Johannes Kepler University Linz, Altenberger Straße 69, 4040 Linz, Austria% \textbackslash\textbackslash
}%

\author{Eva Schöll
\textsuperscript{\S}
}
%\thanks{ Corresponding author: eva.schoell@jku.at}
%\email {eva.schoell@jku.at}
\affiliation{%
 Institute of Semiconductor and Solid State Physics, Johannes Kepler University Linz, Altenberger Straße 69, 4040 Linz, Austria% \textbackslash\textbackslash
}%
\affiliation{Institute for Integrated Circuits and Quantum Computing, Johannes Kepler University, Altenberger Straße 69, 4040 Linz, Austria
}

\author{Armando Rastelli
%\textsuperscript{\P}
}
\thanks{Corresponding authors: ievgen.btytavskyi@jku.at, eva.schoell@jku.at, armando.rastelli@jku.at}
%\email {armando.rastelli@jku.at}
\affiliation{%
 Institute of Semiconductor and Solid State Physics, Johannes Kepler University Linz, Altenberger Straße 69, 4040 Linz, Austria% \textbackslash\textbackslash
}%
%\date{\today}% It is always \today, today,

\input{0_abstract}

\maketitle

%armando.rastelli@jku.at,
%ievgen.brytavskyi@jku.at

%\begin{comment}
\input{1_introduction}
\input{2_design_fabrication}
\input{3_optical_characterization}
\input{4_discussion}
%\end{comment}
%\input{maintext}

\input{5_methods}

\begin{acknowledgments}
We acknowledge Y. Karli for help with the closed-cycle and spectral filtering setup, S.F. Covre da Silva for support with sample growth, A. Halilovic, U. Kainz, S. Brauer, F. Binder, E. Vorhauer for technical support. This project has received funding from the Austrian Science Fund FWF via the Research Group FG5 (10.55776/FG5), the FWF 42 through [F7113] (BeyondC) and from the cluster of excellence quantA [10.55776/COE1] as well as the EU HE EIC Pathfinder challenges action under grant agreement No. 101115575, from the QuantERA II program that has received funding from the European Union’s Horizon 2020 research and innovation program under Grant Agreement No. 101017733 via the projects QD-E-QKD and MEEDGARD (FFG Grants No. 891366 and 906046), Grant Agreement No. 101135288 via the project EPIQUE, and the Johannes Kepler University Linz, Linz Institute of Technology (LIT), via the LIT project PEPSI (LIT-2025-14-YOU-121), and the LIT Secure and Correct Systems Lab, supported by the State of Upper Austria and the Austrian Federal Ministry of Education, Science and Research. The authors also acknowledge support from MUR (Ministero dell’Universita e della Ricerca) through the PNRR MUR project PE0000023-NQSTI. CS was funded by the Austrian Science Fund (FWF) 10.55776/J4784. THL, SH and QB acknowledge funding through the BMFTR funded project QD-E-QKD (FKZ: 16KIS1672K) and QR.N (FKZ: 16KIS2209). THL acknowledges funding through the BMFTR funded project QECS (FKZ: 13N16272).
\end{acknowledgments}

%\section*{Author contributions}
%A.G. and M.P. grew the samples. T.O., J.R., S.M. and Q.B. performed the simulations. I.B., T.O., C.W., T.S., M.P., Q.B., T.M.K. and M.R. carried out device fabrication and characterisation measurements. C.W., M.A., G.U., E.S. and T.O. performed the quantum optical measurements. M.R., S.C., S.H., T.H.L., R.T., T.M.K., E.S. and A.R. supervised the project. E.S., C.W. and T.O., I.B. wrote the manuscript with input from all authors. All authors contributed to discussions and interpretation of the analysis and results.

\bibliography{apssamp}% Produces the bibliography via BibTeX.

\end{document}

%% file: 0_abstract.tex
\begin{abstract}
%\linenumbers % Add this here
\noindent
\begin{comment}

Deterministic sources of entangled photon pairs and indistinguishable photons are expected to play a key role in photonic quantum technologies. Semiconductor quantum dots are promising candidates %for their realisation 
due to their on-demand emission and compatibility with nanophotonic structures. However, current implementations typically face trade-offs between extraction efficiency, Purcell enhancement, spectral bandwith, and charge noise that degrades  indistinguishability and causes blinking. 
Here we demonstrate a quantum dot-based nanophotonic device with electric control combining a high extraction efficiency up to 0.55(6) and Purcell-enhanced decay up to a factor of $\sim$8 via a circular Bragg grating resonator embedded in a p-i-n diode structure. This source is capable of generating wavelength-tunable entangled photon pairs over a range of $\SI{1.6}{nm}$, with raw (corrected) concurrence $> 0.89$ (0.91) and fidelity to the $\ket{\Phi_+}$ state of $>0.94$ (0.95), while suppressing blinking. The same source is also able to emit single, nearly Fourier-transform-limited and highly indistinguishable photons with raw (corrected) $\mathcal{V}_{\text{HOM}} = 0.951(4)$ (0.988(6)). These results show the device's strong potential as a platform for semiconductor-based quantum photonics.
\end{comment}

Deterministic sources of entangled photon pairs and indistinguishable photons are expected to play a key role in photonic quantum technologies. Semiconductor quantum dots are promising candidates due to their on-demand emission and compatibility with nanophotonic structures. However, current implementations face trade-offs between extraction efficiency, Purcell enhancement, as well as charge noise that causes blinking and degrades %photon 
indistinguishability. 
Here we demonstrate a tunable nano-optoelectronic device based on a quantum dot embedded in a p-i-n diode circular-Bragg-grating-resonator and featuring extraction efficiencies up to 0.55(6) and Purcell-factor %enhanced decay up to a factor 
of $\sim$8. The device generates wavelength-tunable entangled photon pairs with suppressed blinking and raw (corrected) concurrence $> 0.89$ (0.91) %and raw fidelity to the $\ket{\Phi_+}$ state of $>0.94$ %(0.95) 
over a range of $\SI{1.6}{nm}$. The very same source also emits single, nearly Fourier-limited and highly indistinguishable photons with raw (corrected) $\mathcal{V}_{\text{HOM}} = 0.951(4)$ (0.988(6)). These results demonstrate a viable platform for semiconductor quantum photonics.

\end{abstract}

%% file: 1_introduction.tex
Photonic quantum technology hinges on the availability of scalable quantum light sources and qubit-photon interfaces, with applications ranging from quantum key distribution (QKD) to large-scale quantum networks~\cite{Gisin2007,Wang2025}, quantum-enhanced imaging and telescopy~\cite{Defienne2024,Czupryniak2023}, and photonic quantum computation~\cite{Xanadu2025,PsiQuantum2025}. Current quantum light sources rely predominantly on non-linear optical processes or attenuated lasers~\cite{Flamini2018, Dutt2024}. The probabilistic nature of these sources imposes a fundamental trade-off between low multi-photon emission and high efficiency. 
Deterministic quantum emitters offer a compelling alternative, enabling on-demand generation of single, indistinguishable and entangled photons~\cite{Senellart2017}.
Among those, epitaxial semiconductor quantum dots (QDs)~\cite{Lodahl2018, Heindel2023, Michler2024} stand out due to their ability to generate highly indistinguishable single photons~\cite{Santori2002, zhai2022quantum}, and entangled photon pairs~\cite{huber2018, Salter2010, Cogan2023clusterstate} with high extraction efficiency~\cite{Ding2025,rota2024}. Their emission properties can be engineered during growth and actively controlled using external fields, enabling the demonstration of quantum network primitives such as quantum teleportation~\cite{Laneve2025, Strobel2025} and entanglement swapping~\cite{beccaceci2025allphotonicentanglementswappingremote} with photons from remote emitters as well as entanglement-based QKD~\cite{Schimpf2021_qkd, BassoBasset_2021_qkd}, spin-photon interfaces~\cite{gao2012,degreve2013, Luo2019, Chan2023}, and quantum registers~\cite{Appel2025}.

The main challenges with QDs stem from their solid state nature. This leads to noise -- originating from lattice, charge, and spin fluctuations -- as well as low extraction efficiency $\eta_{\text{ext}}$ due total internal reflection. 
% to the on-time fraction $\eta_{\text{blink}}$ of $\sim0.3$ for GaAs QDs~\cite{rota2024} under resonant excitation.

Noise is often dominated by charges in the QD surroundings that produces spectral wandering and thus broadened emission lines. Charges can also be captured by the QD, resulting in blinking and reduced source efficiency $\eta_{\text{blink}}$, especially under resonant excitation~\cite{davanco2014, Reigue2018}. Embedding QDs in diode structures~\cite{Warburton2000, Loebl2017} alleviates charge noise in and near them and can lead to Fourier-transform-limited emission lines and suppressed blinking~\cite{Ludwig2017, Schimpf2021c, zhai2020, zhai2022quantum}. Diodes also allow deterministic charge state control, enabling efficient spin-photon interfaces~\cite{zaporski2023, michaels2021multidimensional}, as well as tuning of properties such as biexciton (XX) and exciton (X) recombination energies, and lifetimes, via the quantum-confined Stark effect (QCSE)~\cite{patel2010, Trotta2013, undeutsch2025}. 

$\eta_{\text{ext}}$ can be enhanced by integrating the QDs into nanoscale photonic structures, including waveguiding structures and nanocavities~\cite{Davanco2013, Nowak2014, Lodahl2015, Somaschi2016}. Among those, reflector-backed circular Bragg grating resonators (CBRs) offer a broadband $\eta_{\text{ext}}$ exceeding $80\mathtt{\%}$~\cite{Liu2019,wang2019}, Purcell enhancement ($F_P>25$~\cite{rickert2024}), and near-Gaussian far-field emission. These properties make CBRs highly attractive for bright and fast quantum light sources as well as spin-photon interfaces. However, charge noise is exacerbated in photonic structures due to the QD's proximity to etched and defective surfaces~\cite{Liu2018nanofab,hu2015defect,wang2005,Yang2022}, necessitating the use of diodes.

Despite various proposed designs~\cite{Schall2021, Singh2022, Wijitpatima2024, Buchinger2023, Barbiero2025}, ``diode-CBRs'' bringing together significant Purcell enhancement, high extraction efficiency, and efficient fibre-mode coupling have not yet been experimentally realised. 
In this work, we design, fabricate, and characterise a p-i-n diode-CBR structure deterministically placed around a GaAs QD with emission wavelength around  $\sim\SI{780}{nm}$. This nano-optoelectronic device combines high performance in all the above characteristics. 
%To reach high values for these characteristics for the targeted emitter wavelength range, we develop an optimisation routine that results in a desired combination of the three mentioned emission characteristics. 
Specifically, we demonstrate the efficient generation of pure polarisation-entangled photon pairs with fidelity % after unitary transformation (see methods section~\ref{method:entanglement}) 
of up to $\mathcal{F}_{\ket{\Phi_+}}^{\text{raw}}=0.968(5)$, negligible blinking, $\eta_{\text{ext}}$ of up to 0.55(6) and Purcell factor $F_P$ of up to 8.2. The high emission quality can be maintained over a tuning range of $\pm\,\SI{0.8}{nm}$, making this device suitable for applications involving remote emitters or interfacing with atomic quantum memories~\cite{wolters2017simple}. Thanks to the charge control and broad spectral response of the Purcell enhancement we show that the negative trion in the very same device generates nearly Fourier-transform-limited single photons with near-unity indistinguishability of $V_{\text{HOM}}^{\text{raw}}=0.951(4)$, making diode-CBRs promising candidates for spin-photon interfaces.

%The challenge lies in engineering a device that combines high extraction and Purcell enhancement with a stable, blinking-free environment. This platform enables the generation of highly entangled photon pairs with wide wavelength tunability via the QCSE. 

%The question is therefore: can we achieve near-unity $\eta_{\text{blink}}$ similar to planar diodes~\cite{Schimpf2021c} but with the benefits of enhanced radiative decay rates and extraction efficiency? Furthermore, can we obtain a charge state configuration that is blinking-free over a large voltage range to allow wavelength-tuning via the QCSE? \parTo answer these question,

%suggestion:  The question is therefore: can we realize a tunable, blinking-free source of entangled photons that simultaneously offers high extraction efficiency and Purcell enhancement? Furthermore, can such an integrated platform provide the charge-state control necessary to utilize the negative trion as a source of highly indistinguishable single photons?

%Consequently, the challenge for future entangled photon pair sources is achieving near-unity $\eta_{\text{blink}}$ similar to planar diodes~\cite{Schimpf2021c} but with the benefits of enhanced radiative decay rates and extraction efficiency. Furthermore, can this platform provide the charge-state control necessary to utilize the negative trion as a source of highly indistinguishable single photons?
%

%% file: 2_design_fabrication.tex
\section{Device design and fabrication}
\begin{figure*}[ht]
\includegraphics[width=17cm]{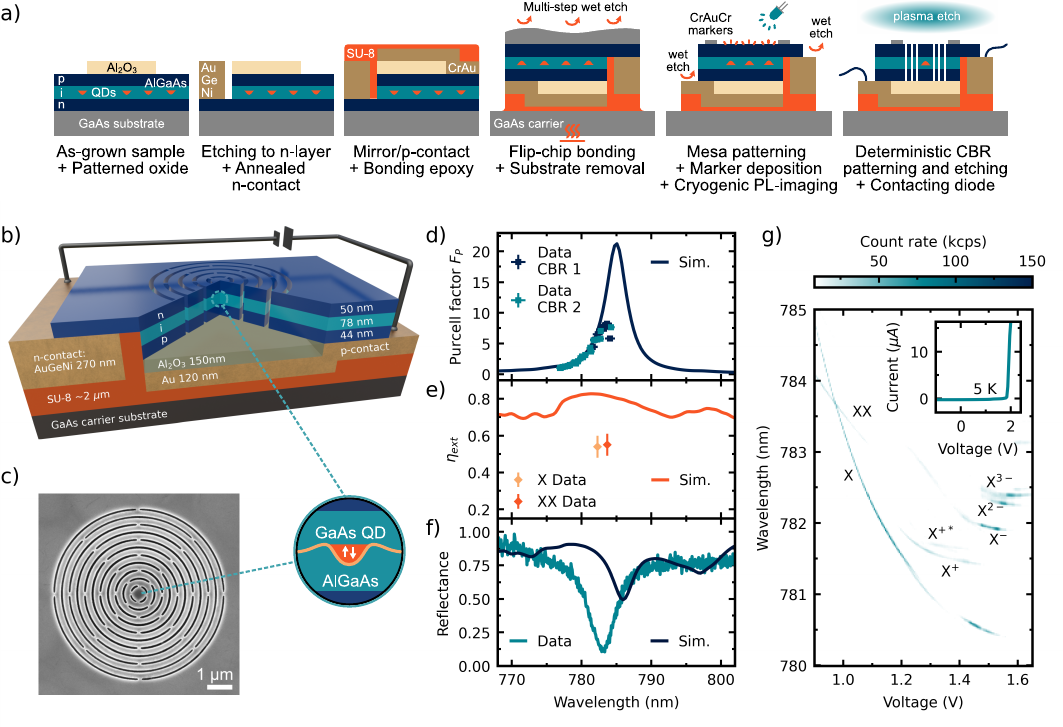}
\caption{\justifying Device structure and basic optical and electrical performance. \textbf{a}, Workflow (from left to right) of deterministic diode-CBR fabrication. \textbf{b}, Sketch of an electrically-contacted CBR containing a GaAs QD integrated in a p-i-n diode on a back reflector. \textbf{c}, Representative SEM image of a fabricated CBR. \textbf{d}, Simulated Purcell factor spectrum with the experimental data obtained from two diode-CBRs. \textbf{e}, Simulated extraction efficiency  $\eta_{\text{ext}}$ spectrum for a numerical aperture of 0.77 and corresponding experimental values for CBR~1. \textbf{f}, Simulated and measured reflectance spectrum of CBR~1 normalised to the reflection on the nearby planar semiconductor membrane. 
\textbf{g}, Colour-coded voltage-dependent photoluminescence spectrum of the QD in CBR~1, excited with a pulsed laser with \SI{770}{nm} wavelength showing different excitonic lines with intensity maxima occurring in different voltage ranges. Inset: Current-voltage characteristics of the p-i-n diode membrane containing the CBRs at \SI{5}{K}.
}
\label{fig:CBRdiode_design}
\end{figure*}

To enable electrical control of a QD in a photonic structure, we place a layer of GaAs QDs obtained by droplet etching epitaxy~\cite{Heyn2009,Gurioli2019} in the intrinsic region of an AlGaAs p-i-n diode membrane~\cite{zhai2020, Schimpf2021c}. The detailed layer sequence is provided in Supplementary Information Section (SI).
%\,\ref{sup:growth} (SI:\,\ref{sup:growth}). 
CBRs deterministically placed with respect to preselected QDs are then used to enhance light-matter interaction and $\eta_{\text{ext}}$ in the detection direction~\cite{Liu2019, wang2019}. To allow electrical connection between the central disk containing the QD and the surrounding areas with contacts, we implement a CBR design with bridges~\cite{Buchinger2023}, specifically optimised for our target wavelength. The main steps of the contacted diode membrane fabrication and deterministic cavity positioning are depicted in Fig.\,\ref{fig:CBRdiode_design}a, details are provided in the methods section\,\ref{method:fab} and SI.
%:\,\ref{sup:fab}. 
Annealing and native oxide removal are employed to obtain ohmic contacts to both p- and n-doped regions. Fig.\,\ref{fig:CBRdiode_design}b shows a sketch of the device, consisting of a diode-CBR placed on a dielectric layer -- here \ce{Al2O3} -- and a bottom \ce{Au} mirror. 

To design the device structure, we carefully consider the interplay between geometrical, electrical, and optical parameters. In particular, the membrane thickness (\SI{172}{nm}) is chosen to ensure reliable electrical connection between contact pads and central disk as well as good optical performance. For the oxide thickness we chose \SI{150}{nm} as for previous CBR implementations without electrical control~\cite{rota2024}. To make sure the doped regions in the bridges are not charge-depleted after etching, we constrain their lateral width to \SI{120}{nm}~\cite{Milanova1999}. Among different designs for the lateral structuring~\cite{Buchinger2023, Barbiero2025} we select the one with a 4-fold ``long way" layout because of the expected performance in the crucial emission characteristics. These are $F_P$, i.e., the ratio of the decay rate in the resonator to the one in bulk, $\eta_{\text{ext}}$, which is defined as the emission efficiency into the upper hemisphere set by the numerical aperture of 0.77 in our simulation, and the overlap of the far-field emitted by a single dipole inside the cavity to a Gaussian mode $\eta_{\text{mode}}$ for efficient coupling into single mode fibres. By using the product of these quantities as figure of merit (FOM), we optimise the geometry for a target QD emission wavelength of \SI{785}{nm} at \SI{4}{K} by the particle swarm method in a 3D-finite-difference-time-domain (FDTD) simulation~\cite{Liu2019} using the central-disk radius, ring periodicity and trench width as parameters. This leads to a central disk radius of \SI{295}{nm}, a ring periodicity of \SI{303}{nm} and a trench width of \SI{88}{nm} as optimal CBR parameters.

Several CBRs with varying central disk radii and the same nominal trench widths and periodicity were fabricated around different QDs on the same diode membrane. A top-view scanning electron microscopy (SEM) image of a representative CBR is shown in Fig.\,\ref{fig:CBRdiode_design}c, with a geometry satisfying the targeted values up to SEM measurement precision. 
With FDTD we calculate also the wavelength-dependent $F_P$ (see Fig.\,\ref{fig:CBRdiode_design}d) and $\eta_{\text{ext}}$ for a NA of 0.77 -- as in our optical setup (see Fig.\,\ref{fig:CBRdiode_design}e). The corresponding simulated performance parameters for the wavelength of \SI{785}{nm} are $F_p=21.2$, $\eta_{\text{ext}}=0.8$ and $\eta_{\text{mode}}=0.62$. The position of the fundamental mode seen in the $F_P$ spectrum can be identified through reflectance spectra, which we use to experimentally calibrate the fabrication process and tune the CBR parameters to the desired target wavelength. Figure\,\ref{fig:CBRdiode_design}f shows the calculated reflectance spectrum together with the experimental data for a CBR-QD system (CBR\,1), which we further investigate in the following. Details on the relative position of the $F_P$ peak with respect to the reflectance minimum~\cite{krieger2024} as well as residual polarisation-splitting of cavity modes are provided in the SI.

%% file: 3_optical_characterization.tex
\section{Quantum-optoelectronic properties}
By applying a voltage $V$ across the diode membrane and measuring the corresponding current, we observe a nearly ideal diode behaviour, as shown in the inset of Fig.\,\ref{fig:CBRdiode_design}g. Negligible current flows in reverse and forward direction up to $V\approx \SI{1.7}{V}$, which is the diode built-in voltage corresponding to the band gap energy of the doped \ce{Al_{0.15}Ga_{0.85}As} layers at low temperatures. The electric field at this voltage is nominally zero and increases with decreasing $V$.
Optical characterisation of the devices is performed using a cryogenic micro-photoluminescence setup (see Methods\,\ref{method:upl}). A pulsed laser with a repetition rate of \SI{80}{MHz}, pulse duration of $\sim\SI{3.5}{ps}$, and $\sim\SI{770}{nm}$ wavelength is used to excite the QD through its dense ladder of excited states. 

Fig.\,\ref{fig:CBRdiode_design}g shows photoluminescence spectra of the QD in CBR\,1 recorded as a function of $V$. The emission lines show a red-shift with decreasing voltage (increasing electric field) due to the QCSE.
Towards higher voltages, the conduction and valence band are only slightly tilted and the electron quasi-Fermi level lies above the lowest-energy confined electron state in the conduction band of the QD material, leading to negative charging. This leads to emission dominated by the twice (X$^{2-}$) and three times (X$^{3-}$) negatively charged states~\cite{Warburton2000}. 
With decreasing $V$, first the negative trion X$^{-}$ becomes visible for $V\approx 1.55-\SI{1.4}{V}$ and then the neutral exciton X for $V\lesssim\SI{1.54}{V}$. Dim signals from the positive trion X$^{+}$ and its excited states~\cite{daSilva2021} are seen for $V\approx 1.46-\SI{1.2}{V}$ and compete with the X emission.
For $V\lesssim\SI{1.2}{V}$, the spectra are eventually dominated by charge neutral configurations, with a weak signal from the biexciton (XX) appearing for $V\lesssim\SI{1.08}{V}$ in addition to X. Simulations (see SI)
%:\,\ref{sup:diode_bands}) 
indicate that for these voltages the electron (hole) quasi-Fermi level lies below (above) the conduction (valence) band edge of the QD material. As a result, charge carrier tunnelling into the QD is suppressed, favouring a stable neutral charge state.

\subsection{Efficient, tuneable, and blinking-free polarisation-entangled photon pair generation}

\begin{figure*}
\includegraphics[width=17cm]{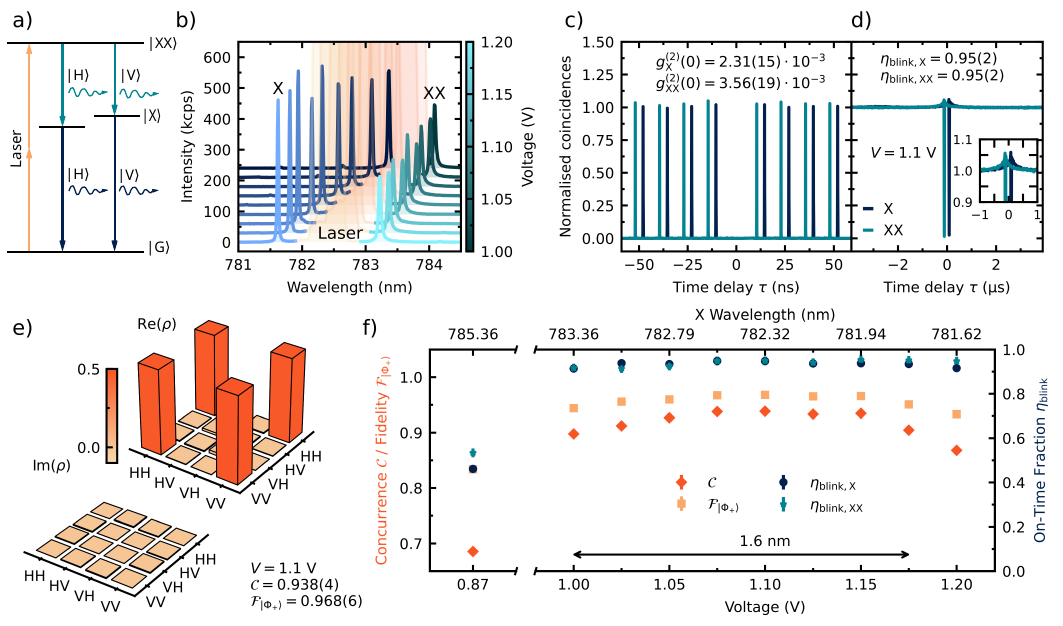}
\caption{\justifying Near blinking-free emission of wavelength tuneable entangled photons under pulsed resonant two-photon excitation. \textbf{a}, Relevant energy levels in a QD. The biexciton state $\ket{\text{XX}}$ is resonantly populated by TPE and recombines via two possible decay paths (H or V polarised) through one of the bright exciton states $\ket{\text{X}}$. Two photons are emitted, first the biexciton (XX, petrol) and subsequently the exciton (X, blue) photon. \textbf{b}, Spectra at different voltages, showing the XX emission at lower energy (higher wavelength) and the X emission at higher energy. With increasing voltage, both lines redshift and the XX relative binding energy $E_b$ decreases. The TPE laser is shown in pale orange for better visibility. \textbf{c}, Second-order autocorrelation histogram of XX and X at $V=\SI{1.1}{V}$ showing clean single-photon emission. \textbf{d}, Same measurement on longer timescales, displaying only a small bunching on a sub-microsecond timescale, corresponding to residual blinking and an on-time fraction $\eta_{\text{blink}}=0.95(2)$. Time axes are shifted for clarity. \textbf{e}, Reconstructed real and imaginary part of the two-photon polarisation density matrix at \SI{1.1}{V}. The corresponding concurrence $\mathcal{C}$ and fidelity $\mathcal{F}_{\ket{\Phi_+}}$ are indicated. \textbf{f},$\eta_{\text{blink}}$ and entanglement degree as a function of $V$. The device shows almost blinking-free behaviour with $\eta_{\text{blink}}>90\%$ and high entanglement with $\mathcal{F}_{\ket{\Phi_+}}>0.95$ and $\mathcal{C}>0.898$ over an X tuning range of \SI{1.6}{nm}.}
\label{fig:cascade}
\end{figure*}

To investigate the performance of our device with respect to entanglement generation, we use the biexciton-exciton cascade. To this end, we resonantly populate the biexciton state $(\ket{\text{XX}})$ via two-photon excitation (TPE) using a laser pulse duration of $\sim\SI{3.5}{ps}$, as illustrated in the energy level scheme in Fig.\,\ref{fig:cascade}a, and reach a preparation fidelity of $\eta_{\text{prep}}=0.90(2)$, see SI.
%:\,\ref{sup:extr_eff}. 
After excitation, the system recombines to the ground state within a characteristic time $T=T_{1}^{\text{XX}}+T_{1}^{\text{X}}$, the sum of the $\ket{\text{XX}}$ and $\ket{\text{X}}$ lifetimes, via one of two bright exciton states $(\ket{\text{X}})$ through the emission of two photons, a XX and an X photon with an energy difference $E_b$ (relative biexciton binding energy). If the two $\ket{\text{X}}$ levels are degenerate, the two possible decay paths are indistinguishable, and the polarisation of the emitted XX-X pair is ideally in the maximally entangled $\ket{\Phi_+}$ Bell state, a process that has been widely explored~\cite{muller2014, huber2018}, also with QDs in CBRs~\cite{Liu2019, wang2019, rota2024}.

To study the generation of entangled photon pairs we focus on the QD in CBR\,1 because of its almost degenerate excitonic levels (fine structure splitting, FSS, of up to $\SI{1.1(1)}{\mu eV}$) and nearly unpolarised emission ($<$14(1)\%, see SI.
%:\,\ref{sup:chap_DOP} and \ref{sup:chap_FSS}).
Spectra of this QD under $\pi$-pulse TPE are shown in Fig.\,\ref{fig:cascade}b for different values of $V$. 
Due to the QCSE, both XX and X transitions redshift with increasing field (decreasing $V$), however the exciton exhibits a higher polarisability, leading to a stronger tuning. For the voltage range from 1.2 to \SI{1.0}{V}, the XX (X) transition shifts by \SI{1.7}{meV} (\SI{3.5}{meV}). Consequently, $E_b$ decreases with decreasing $V$, and the lines cross at approximately \SI{0.93}{V}, as previously observed for both GaAs and InGaAs QDs~\cite{undeutsch2025, Trotta2013}.

To verify the deterministic nature of the emission process and quantify the impact of blinking, we measure the second-order autocorrelation function $g^{(2)}(\tau)$ with a Hanbury Brown and Twiss interferometer. On short timescales in the order of few repetition cycles of the excitation laser, we observe a pronounced anti-bunching around $\tau=0$, demonstrating the clean single-photon nature of both transitions with $g^{(2)}_{\text{XX}}(0)=3.56(19)\cdot10^{-3}$ and $g^{(2)}_{\text{X}}(0)=2.31(15)\cdot10^{-3}$, as shown in Fig.\,\ref{fig:cascade}c for $V=\SI{1.1}{V}$. We attribute the residual multi-photon emission to minor contributions from re-excitation~\cite{hanschke2018} and unfiltered laser leaking into the signal. 
On longer timescales, $g^{(2)}(\tau)$ reveals a small bunching on the timescale of sub-microseconds for both transitions (see Fig.\,\ref{fig:cascade}d), which we ascribe to blinking~\cite{Jahn2015}. The ratio of the uncorrelated coincidences at long time delay to the bunching peak amplitude yields a near unity on-time fraction $\eta_{\text{blink}}=0.95(2)$ via fitting (see methods\,\ref{method:g2}). This value represents an improvement of $\eta_{\text{blink}}$ by a factor $\sim$3 compared to non-diode CBRs with similar QDs~\cite{rota2024} and demonstrates that the QD is consistently in a charge neutral state in spite of the photonic cavity processing.
We attribute the remaining blinking to charge carriers, which are generated by the laser -- possibly by photoionisation of charge traps at the surface -- and get captured in the QD. 

Source efficiency and Purcell enhancement are key characteristics of photonic structures, which we now characterise for CBR\,1 (see SI for more details).
%:\,\ref{sup:purcell} and \ref{sup:extr_eff} for more details). 
Lifetime measurements at $\SI{1.1}{V}$ yield $T_{1}^{\text{XX}}=\SI{15(1)}{ps}$ and $T_{1}^{\text{X}}=\SI{40(1)}{ps}$ (see SI).
%:\ref{supp:cbr1_T1}). 
These values are 5-8 times shorter than the typical lifetimes observed for similar GaAs QDs in absence of a microcavity~\cite{huber2018}, suggesting substantial light-matter-interaction enhancement. To estimate $F_P$ we must take into account the presence of an electric field, which results in increased lifetimes due to a decreased electron and hole wave-function overlap~\cite{undeutsch2025}. 
Comparing the measured $T_{1}^{\text{XX}}$ and $T_{1}^{\text{X}}$ with the intrinsic lifetimes without resonator under the same estimated electric field and assuming negligible non-radiative recombination rates yields an estimated maximum Purcell factor of $F_P^{\text{max}}=8.2$ for XX and 7.9 for X (see SI).
%:\ref{sup:purcell}). 
By performing measurements at different electrically-controlled detunings from the cavity mode of CBR\,1 and 2, we obtain the data points shown in Fig.\,\ref{fig:CBRdiode_design}d. We see that the wavelength dependence of the estimated $F_P$ follow a Lorentzian curve with a Q-factor of $\sim 156$. For comparison, the simulation yields $F_P^{\text{max}}\sim21.1$ and Q$\sim215$.

To quantify $\eta_{\text{ext}}$ for a lens with a $NA=0.77$, we characterise the transmission losses of our optical setup and record the single photon counts on our detectors. We obtain $\eta_{\text{ext}}=$0.55(6) (0.54(6)) for XX (X) photons, resulting in a coincidence rate of \SI{196}{kcps} on the superconducting-nanowire single-photon detectors and a probability of 0.25(7) to collect a photon pair in front of the first lens.
The measured $\eta_{\text{ext}}$ is to our knowledge the highest so far for a diode structure with GaAs QDs, but lies below the simulation predictions of $\sim0.8$, see  Fig.\,\ref{fig:CBRdiode_design}e. 

We qualitatively attribute the observed deviations between measured and simulated $F_P^{\text{max}}$, Q-factor and $\eta_{\text{ext}}$ to scattering and absorption losses that were not accounted for in the simulations, as well as fabrication imperfections like asymmetries in the structure or displacement of the QD inside the cavity, which additionally leads to a polarised emission~\cite{peniakov2024}.

To measure the degree of polarisation-entanglement between the XX and X photons, we perform a full quantum-state tomography and reconstruct the two-photon density matrix with a maximum likelihood method~\cite{james2001} for different values of $V$ (see methods\,\ref{method:entanglement}). As an example, Fig.\,\ref{fig:cascade}e shows the real and imaginary part of the density matrix at $\SI{1.1}{V}$. We extract a concurrence of $\mathcal{C}=0.938(4)$ and, upon a unitary transformation (see methods\,\ref{method:entanglement}), a fidelity to the $\ket{\Phi+}$ Bell state of $\mathcal{F}_{\ket{\Phi_+}}=0.968(5)$. 
After accounting for multi-photon contributions~\cite{huber2018, fognini2019}, the concurrence increases to $\mathcal{C}^{\text{corr}}=0.943(4)$ and the fidelity to $\mathcal{F}_{\ket{\Phi_+}}^{\text{corr}}=0.972(6)$. 
These values are comparable to the state of the art for GaAs QDs~\cite{huber2018, rota2024}. Most importantly, due to the diode structure, the QCSE can be exploited to tune the wavelength of the XX and X photons. 
Figure\,\ref{fig:cascade}f shows the X emission wavelength, on-time fraction, entanglement fidelity and concurrence as a function of the applied voltage. Varying $V$ from 1.0 to \SI{1.175}{V} results in a tuning range of \SI{1.6}{nm} (\SI{3.2}{meV}) for the X transition. Across this range, both transitions remain nearly blinking-free with $\eta_{\text{blink}}>0.9$. Simultaneously, the degree of entanglement remains high with $\mathcal{C}>0.89$ and $\mathcal{F}_{\ket{\Phi_+}}>0.94$ (see SI).
%:\,\ref{sup_chap_Entanglement}). 
Taking into account the measured $g^{(2)}(0)$, lifetimes, and FSS yields an intrinsic upper bound of $\mathcal{C}\leq0.98 $~\cite{hudson2007, rota2024}. In particular, the short lifetimes mitigates the degrading effect of the FSS. We therefore attribute the remaining entanglement imperfection primarily to the AC-Stark effect~\cite{seidelmann2022}, with a possible minor contribution from residual emission polarisation in CBRs~\cite{laneve2025wavevector}. 
For $V\gtrsim$\SI{1.2}{V}, the XX and X intensity drops and additional lines appear due to competing charge configurations, preventing clean population of the biexciton state. For $V\lesssim$\SI{1.0}{V}, the biexciton binding energy drops below \SI{1.4}{meV}, making laser filtering challenging. Nevertheless, we are able to excite the biexciton at $V=\SI{0.87}{V}$, after the XX-X crossing, where the biexciton is ``antibinding'' with a $E_b\simeq$\SI{-2.2}{meV}. Although in this configuration the laser is blue-detuned from the exciton transition and can unintentionally populate the exciton also via a phonon-assisted process~\cite{juska2020}, the two-photon state still exhibits a high degree of entanglement with $\mathcal{C}=0.686(4)$ and $\mathcal{F}_{\ket{\Phi_+}}=0.834(6)$.

\subsection{Highly indistinguishable and Fourier-transform-limited single photons}
\begin{figure*}
\includegraphics[width=17cm]{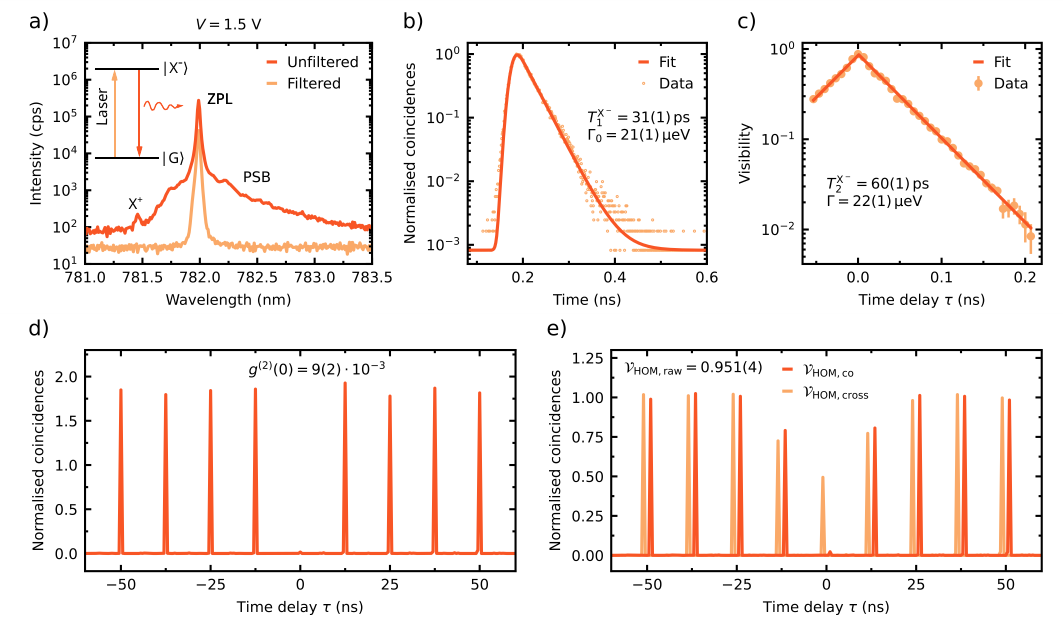}
\caption{\justifying Indistinguishable, nearly Fourier-transform-limited single photons under $\pi$-pulse resonant excitation at $V=\SI{1.5}{V}$. \textbf{a}, Semi-logarithmic unfiltered and filtered resonance fluorescence spectra of the negatively charged trion X$^-$. Inset: Two-level quantum system. \textbf{b}, Lifetime measurement with fit, yielding $T_{1}^{\text{X}^-}=\SI{31(1)}{ps}$. \textbf{c}, Semi-logarithmic first-order autocorrelation function $|g^{(1)}(\tau)|$ and Lorentzian fit, yielding a linewidth of $\Gamma=\SI{22(1)}{\mu eV}$, a factor $1.05^{+0.07}_{-0.05}$ above the Fourier-transform-limit. \textbf{d}, Second-order autocorrelation histogram, yielding $\Tilde{g}^{(2)}_{\text{X}^-}(0)=9(2)\cdot10^{-3}$. \textbf{e}, Histograms of Hong-Ou-Mandel interference of co- and cross polarised photons, yielding a raw (corrected) visibility of $\mathcal{V}_{\text{HOM}}=0.951(4)$ (0.988(6)).}
\label{fig:trion}
\end{figure*}

Another important requirement for quantum light sources is photon indistinguishability. Since the radiative cascade following the biexciton excitation inherently limits the indistinguishability of the emitted photons due to the non-separability of the two-photon output state~\cite{simon2005, Huber2013, Scholl2020}, we choose a clean two-level system (see inset Fig.\,\ref{fig:trion}a) in the same QD discussed above to assess the capability of our device to generate indistinguishable photons. 
Specifically, we select the negatively charged trion X$^-$ also because of its appeal for spin-photon interfaces~\cite{michaels2021multidimensional} and address it by resonance fluorescence in a cross-polarisation configuration~\cite{kuhlmann2013} with a laser suppression of $\sim 10^4$ in spite of the nanostructured device.

Figure\,\ref{fig:trion}a shows the X$^-$ spectrum in the centre of the charge plateau at a voltage of $\SI{1.5}{V}$ and for an average laser power of \SI{2}{nW}, corresponding to $\pi$-pulse excitation.
The controlled charge environment, enabled by the diode structure together with the resonant excitation, results in a clean spectrum with only one prominent emission line stemming from the zero phonon line (ZPL) of the respective transition and small contributions from the phonon sideband (PSB), the positive trion (X$^+$) and excitation laser, as shown in Fig.\,\ref{fig:trion}a. 

To assess how close the X$^-$ is to an ideal two-level quantum emitter, we measure the lifetime $T_{1}^{\text{X}^-}$, second-order $g^{(2)}(\tau)$, and first-order correlation $|g^{(1)}(\tau)|$ autocorrelation functions of the emitted photons. 
The time trace of $\text{X}^-$ photon counts following pulsed excitation is shown in Fig.\,\ref{fig:trion}b, along with a fit (see methods\,\ref{method:lifetime}), yielding $T_{1}^{\text{X}^-}=\SI{31(1)}{ps}$. Comparing this value with the typical lifetime of a negative trion of $\sim\SI{239(4)}{ps}$ in a GaAs QD in bulk reveals a Purcell factor of $F_P\approx7.7(3)$ (see SI).
%:\ref{sup:chap_TRION}).
From the measured $T_{1}^{\text{X}^-}$, the Fourier-transform-limited linewidth is $\Gamma_0=\hbar / T_{1}^{\text{X}^-}=\SI{21(1)}{\mu eV}\equiv\SI{5.1(2)}{GHz}$. To determine the actual linewidth of the transition, we measure the $|g^{(1)}(\tau)|$ with a Michelson interferometer. The data, together with a Lorentzian fit, are shown in Fig.\,\ref{fig:trion}c in a semi-logarithmic plot. The linear behaviour confirms that the line shape is predominantly Lorentzian. From the fit, we extract a coherence time of \SI{60(1)}{ps} corresponding to a linewidth of $\Gamma=\SI{22(1)}{\mu eV}$. This value is only a factor of $1.05^{+0.07}_{-0.05}$ above the Fourier-transform-limit. To the best of our knowledge, this is the best value reported to date for QDs embedded in photonic structures~\cite{pedersen2020}, reaching values comparable to those obtained in planar structures~\cite{zhai2020, zhai2022quantum}.

For all subsequent measurements, we filter remaining scattered laser light with an additional volume Bragg grating filter (see SI).
%:\,\ref{sup:chap_TRION}), 
resulting in the spectrum shown in Fig.\,\ref{fig:trion}a. The second-order autocorrelation function $g^{(2)}(\tau)$ shown in Fig.\,\ref{fig:trion}d exhibits clean anti-bunching, confirming the single-photon nature of the emission with a multi-photon emission probability of $\Tilde{g}^{(2)}(0)=9(2)\cdot10^{-3}$. Analysing the data on long time scales reveals blinking on timescales of the order of tens of microseconds with an on-time fraction of $\eta_{\text{blink}}=0.57(2)$ (see SI)
%:\,\ref{sup:chap_TRION})
, which leads to $g^{(2)}(0)=\Tilde{g}^{(2)}(0)/\eta_{\text{blink}}$. The histogram is normalised accordingly. We attribute the significantly more pronounced blinking compared to that observed for the neutral configuration (cf. previous section) to the lower electric field at which the measurements on X$^-$ are performed. As a consequence, we argue that the probability that carriers produced by trap photoionisation are captured by a QD increases, leading to charge state fluctuations. The coexistence of other excitonic states, such as the positively charged trion, as visible in the voltage-dependent photoluminescence spectra in Fig.\,\ref{fig:CBRdiode_design}g as well as the increased $\eta_{\text{blink}}=0.87(2)$ at lower excitation intensity (see SI)
%:\,\ref{sup:chap_TRION}) 
support this scenario. 

Finally, to quantify the indistinguishability of the emitted photons, we determine the Hong-Ou-Mandel (HOM) interference visibility through measuring the two-photon interference in an unbalanced Mach-Zehnder interferometer (see methods\,\ref{method:HOM}). Here, the additional volume Bragg grating filter suppresses only the distinguishable photons from the PSB. The corresponding histogram for the co- and cross-polarised measurement is shown in Fig.\,\ref{fig:trion}e. From that, we extract a raw (corrected) visibility of $\mathcal{V}_{\text{HOM}}=0.951(4)$ (0.988(6)), indicating near-unity indistinguishability for photons separated by $\sim\SI{12.5}{ns}$. 
We emphasize that the HOM measurement only assesses the indistinguishability of subsequently emitted photons, and does not capture slow processes like spectral wandering. The Fourier-transform limited linewidth reported above, determined via a time-averaged measurement, is a much more stringent test of optical quality and implies similarly high indistinguishability for photons separated on macroscopic timescales or coming from independent emitters.

%We determine a small intrinsic fine structure splitting of $\SI{1.5}{\mu eV}$, which leads to a temporal oscillation of the output state between the $\ket{\Phi_+}$ and $\ket{\Phi -}$ Bell states, which leads to a degradation in the time-averaged fidelity to the $\ket{\Phi_+}$ state. Here, this oscillation is slow compared to the lifetimes, and no external tuning method is required to eliminate the fine structure splitting~\cite{trotta2015, rota2024}.
%However, if the lifetimes approach the excitation pulse duration, the AC Stark effect degrades the degree of entanglement~\cite{seidelmann2022, bassobasset2023}. This effect can be mitigated by employing shorter laser pulses. However, $E_B$ is only \SI{2.7}{meV} at \SI{1.1}{V}, which imposes a limitation on the possible pulse duration of \SI{3.5}{ps}, where the laser is spectrally still isolated from the emission lines. To improve the laser suppression, we place two volumetric Bragg grating filters in the excitation path tuned in a way, that it cuts out a narrow spectral part at the two emission wavelengths.

%% file: 4_discussion.tex
\section{Conclusions}

In summary, we present an efficient and tuneable entangled photon pair source based on a GaAs QD deterministically placed in a diode-CBR, combining several key characteristics in a single device. The source exhibits negligible blinking, with an on-time fraction of $\eta_{\text{blink}}>0.9$, a photon extraction efficiency of up to 0.55(6), and a high degree of entanglement with fidelity (concurrence) $\mathcal{F}^{\text{raw}}_{\ket{\Phi_+}}>0.94$ ($\mathcal{C}^{\text{raw}}>0.89$), maintained across a wide exciton tuning range of \SI{1.6}{nm}. 
This spans $\sim21\%$ of the emission wavelengths of the QD ensemble~\cite{daSilva2021}, making the tuning range of the device well-suited for quantum network primitives involving remote emitters and/or elements with predefined operation wavelengths, such as quantum memories based on atomic clouds (e.q. $^{87}$Rb D$_1$ and D$_2$ lines) or lasers with fixed wavelengths. 

The Purcell enhancement of up to $\sim$8 shortens the emitter lifetime, which has two important consequences for entangled photon generation. First, it renders the small intrinsic fine structure splitting of GaAs QDs~\cite{daSilva2021} negligible: although such a splitting would ordinarily degrade the time-averaged fidelity to the $\ket{\Phi_+}$ state, the short lifetimes, corresponding to Fourier-transform-limited linewidths of $>\SI{15}{\mu eV}$ for the exciton alleviates this effect, eliminating the need for any external field tuning. Second, the short average duration of the radiative cascade ($T\sim\SI{55}{ps}$) potentially supports GHz-rate operation, potentially increasing coincidence rates by more than an order of magnitude compared to the current demonstration. At the same time, as the XX lifetime approaches the excitation pulse duration, the AC Stark effect begins to limit the degree of entanglement~\cite{seidelmann2022,bassobasset2023}, while the finite relative XX binding energy prevents arbitrarily short pulses as a remedy. Excitation schemes using far-detuned lasers offer a promising path to overcoming this trade-off~\cite{yan2025, piccinini2025}.

A central remaining challenge is achieving highly entangled pairs of indistinguishable photons simultaneously, as the indistinguishability of cascaded photons is limited by the intrinsic lifetime ratio of the biexciton-exciton cascade~\cite{Scholl2020}. Differential Purcell enhancement~\cite{baltisberger2025,behrends2026} and operation via the QCSE~\cite{undeutsch2025} are complementary strategies to address this limitation, and the diode-CBR geometry is well suited to realising both in tandem. Residual linewidth broadening and blinking can possibly be further reduced through surface passivation~\cite{zhao2025} and increased material quality. 
Achieving these properties simultaneously will be an important step forward toward semiconductor-based quantum photonic technologies, requiring indistinguishable photons from remote sources, such as most notably entanglement swapping in quantum repeater networks. But even in its current form, our device is well suited for near-term quantum communication applications such as entanglement-based quantum key distribution.

In addition, we demonstrate that the negative trion in the very same QD can be used as a nearly ideal single-photon emitter with close to Fourier-transform-limited linewidth ($\Gamma/\Gamma_0=1.05^{+0.07}_{-0.05}$), high indistinguishability ($\mathcal{V}_{\text{HOM}}^{\text{raw}}=0.951(4)$), and high extraction efficiency, making our diode-CBR very suitable for spin–photon interfaces and generation of photonic cluster states \cite{zaporski2023, michaels2021multidimensional,schimpf2025optical}. To date, comparable performance has only been approached using a more complex open-cavity architectures~\cite{tomm2021}. Our approach, by contrast, provides these capabilities in a more practical and integrated platform, opening an alternative pathway toward scalable quantum networks and photonic quantum information processing.

%% file: 5_methods.tex
\cleardoublepage
\setcounter{subsection}{0}
\section{Methods}
\subsection{Simulation}\label{method:simulation}
The simulations for the CBR design were performed with the 3D FDTD solver from Ansys Lumerical. The QD is modeled as a dipole oriented along the x-axis and placed at the centre of the photonic cavity. A near-field monitor was placed above the structure to extract the emission characteristics. More details are provided in SI:\ref{chap:device_design}.

\subsection{Sample fabrication}\label{method:fab}
The heterostructure used for this work consists of a sequence of a thin GaAs cap layer, an n-doped \ce{Al_{0.15}Ga_{0.85}As} layer, an undoped \ce{Al_{0.15}Ga_{0.85}As}--\ce{Al_{0.33}Ga_{0.67}As} bilayer, a layer with GaAs QDs, an \ce{Al_{0.33}Ga_{0.67}As} barrier and a p-doped \ce{Al_{0.15}Ga_{0.85}As}--GaAs bilayer, forming a n-i-p diode with embedded GaAs QDs at its center. The total heterostructure thickness is \SI{172}{nm} and the heterostructure is grown over an Al-rich sacrificial layer for selective etching during processing; a more detailed description is given in SI:\ref{sup:growth}. After depositing and patterning a \SI{150}{nm} thick \ce{Al2O3} oxide spacer, ohmic contacts are formed on the buried n-doped layer and a \SI{120}{nm} thick Au layer is deposited. This layer plays the double role of a broadband backside reflector and p-contact where the metal is in touch with the p-doped GaAs layer. Then the sample is bonded with SU-8 photoresist to a GaAs carrier, the original substrate is removed using a wet etching process, and the remaining membrane is patterned into separate mesa devices with individual contacts. Subsequently, a grid of metal markers is formed on the surface of the membrane, enabling QD position mapping with micro-photoluminescence imaging~\cite{sapienza2015, rota2024}, and CBRs are fabricated around selected QDs by using electron beam lithography and inductive coupled plasma reactive ion etching, see \ref{fig:CBRdiode_design}a and SI:\ref{sup:fab}.

\subsection{Micro-photoluminescence setup}\label{method:upl}
The sample is placed in a closed-cycle cryostat and cooled to \SI{4.5}{K}. The excitation is performed with a tunable pulsed Ti:Sa laser with a repetition rate of \SI{80}{MHz} and pulse duration of \SI{108}{fs}. A 4f-pulse shaper is employed to spectrally cut the laser pulses, resulting in pulse durations of $\sim\SI{3.5}{ps}$ for TPE and $\sim\SI{2.7}{ps}$ for direct resonant excitation, while also enabling fine tuning of the excitation wavelength. For TPE of the biexciton we use a 90:10 beamsplitter to combine excitation and collection. For resonance fluorescence of the negative trion, we use a cross-polarisation configuration. The photoluminescence signal is collected with an aspheric lens (NA=0.77) inside the cryostat sample-chamber and subsequently coupled into a single-mode fibre. Thus the signal can flexibly be sent to either a spectrometer equipped with a charge-coupled-device (CCD) camera or different analysis setups (see below). Further details on the setup are given in Ref.~\cite{undeutsch2025}.

\subsection{Linewidth measurement}\label{method:michelson}
To characterise the coherence and thereby the linewidth of the emitter, we use Michelson interferometry as described in \cite{undeutsch2025}. The signal is split in a free-space 50:50 beamsplitter and sent into the two interferometer beams, one with variable length, both containing a retroreflector. Back on the beam splitter, the two beams interfere and one output is sent through a multimode fibre to the spectrometer. For each time delay, e.g. path length difference between the two beams, we determine the interference fringe visibility, through modifying the phase with a linear piezo stage in one arm. To analyse the data, the resulting intensity on the spectrometer CCD is integrated in a spectral window of $\sim 5$ FWHM around the zero-phonon line. We extract the coherence time $T_2$ by fitting the visibility with a Fourier-transform of a Lorentzian and extract a linewidth $\Gamma=2\hbar/T_2$. 

\subsection{Fine structure splitting and degree of polarisation measurement}\label{subsec:FSS} 
We measure the fine structure splitting (FSS) and degree of polarisation of the QD emission by placing a half-wave plate followed by a fixed polariser in the collection path. The AC-Stark effect effectively imposes a time-dependent additional FSS on the exciton~\cite{seidelmann2022}. To minimise this effect on the measurement of the intrinsic FSS, we excite the $\ket{XX}$ via TPE with short ($\sim$ \SI{3.5}{ps}) pulses at low excitation power ($<~0.2~\pi$-pulse). 
Polarisation-resolved spectra of the emitted photons are acquired with the spectrometer using a \SI{1800}{g/mm} grating by changing the half-wave plate axis. The energy difference between the X and XX peak positions (obtained via Lorenz fits) is then fit with a cosine function, where half of the extracted amplitude corresponds to the FSS. For the degree of polarisation (DOP) $\pi$-pulse excitation is employed. To determine the DOP, we fit the integrated peak intensities as a function of the polarisation angle with a cosine function.

\subsection{Reflectance measurements}\label{subsec:Reflectance}
A spectrally broad, unpolarised halogen light source coupled into a single-mode fibre is used for illumination. In the collection path, a half-wave plate followed by a fixed polariser enables polarisation-resolved detection. The reflected signal is coupled into a single-mode fibre and analysed with a spectrometer equipped with a 300~g/mm grating. To obtain the relative reflectance of the CBRs, the measured reflectance spectrum of each CBR is normalised to a reference spectrum recorded a few micrometers away from the structure on a planar part of the sample.

\subsection{Correlation spectroscopy and filtering}
For correlation spectroscopy measurements, the emission line is filtered using a volume Bragg grating (VBG)–based filtering setup~\cite{undeutsch2025}. It is used to suppress scattered laser and the phonon side band, thereby isolating the zero phonon line. For entanglement measurements, two such filtering setup are employed to additionally separate the biexciton and exciton emission lines. The filtered signal is sent to the respective setup (see below) and then detected by superconducting-nanowire single-photon detectors (SNSPDs) with a timing jitter of \SI{13}{ps}. The detection events are correlated in a time tagger device with a timing jitter of \SI{4}{ps}.

\subsubsection*{Lifetime measurements} \label{method:lifetime}
To measure the excitonic decay dynamics after optical excitation and quantify the lifetime of the excitonic states, we perform cross-correlation measurements between the arrival times of photons from the excitation laser and the photoluminescence signal. The setup instrument response function is determined by performing an auto-correlation measurement of the excitation laser and fitting the resulting data with a Gaussian function. We then fit the time-resolved photon counts -- see, e.g. Fig.\,\ref{fig:trion}(b), with a convolution of the setup instrument response function and a mono-exponential decay for the biexciton and trion, or a bi-exponential decay for the exciton.

\subsubsection*{Second-order correlation measurements}\label{method:g2}
To characterise the single-photon purity and blinking of a transition, we measure the second-order correlation function with a Hanbury Brown and Twiss setup. The filtered signal is directed onto a fibre-based 50:50 beam splitter, with the two outputs connected to SNSPDs. A time tagger records and correlates the detection events and generates a coincidence histogram between the two detectors. For the single-photon purity the events are correlated with a resolution of \SI{10}{ps}. The multi-photon emission probability is given by $\Tilde{g}^{(2)}(0)=A_{\text{centre}}/\overline{A}_{\text{side}}$, with the areas $A_i$ of the centre and side peaks. To determine the peak areas, we integrate all coincidences in a time window of \SI{2}{ns}, much larger than the emitters lifetimes.
The data is normalised by dividing the coincidences by the expectation value for a Poissonian source $R_{\text{1}}R_{\text{2}}t_{\text{bin}}T$, where $R_{\text{1,2}}$ are the count rates on the detectors, $t_{\text{bin}}$ is the time-bin corresponding to \SI{12.5}{ns}, and $T$ is the integration time. For the blinking analysis, the detection events are correlated with a resolution of $\sim\SI{12.5}{ns}$, matching the repetition rate of the laser. The bunching feature is fitted with a double-exponential decay
\begin{equation*}
    g^{(2)}(\tau)=1+A_1\exp{\left(-\left|\frac{\tau}{\tau_1}\right|\right)}+A_2\exp{\left(-\left|\frac{\tau}{\tau_2}\right|\right)}, 
\end{equation*}
with the exponential decay times $\tau_{1,2}$ giving the time-scales of the blinking and the corresponding amplitudes $A_{1,2}$. The on-time fraction is given by $\eta_{\text{blink}}=\frac{1}{1+A_1+A_2}$.

\subsubsection*{Hong-Ou-Mandel (HOM) measurement}\label{method:HOM}
To quantify the indistinguishability, two-photon interference visibility is measured in  a Hong-Ou-Mandel (HOM) experiment. The filtered photons are sent into an unbalanced Mach-Zehnder Interferometer with a path-delay of $\sim\SI{12.5}{ns}$, matching the repetition rate of the laser, so that subsequently emitted photons can interfere in a fibre-based 50:50 beam splitter. The beam splitter outputs are connected to a SNSPD each and the detection events are correlated like above. The HOM visibility is given by $\mathcal{V}_{\text{HOM,raw}}=1-N_{\text{co}}/N_{\text{cross}}$. Here, $N_{\text{co,cross}}$ denotes the relative area of the centre peak for the co-polarised (indistinguishable) and cross-polarised (distinguishable) configuration. They are calculated as described above, with the modification that only peaks at $|\tau|\geq\SI{25}{ns}$ are used for the normalisation. In order to gain access to the indistinguishability we correct $\mathcal{V}_{\text{HOM,raw}}$ via \cite{Santori2002,zhai2022quantum}:

\begin{equation}
    \mathcal{V}_{\text{HOM,cor}}=\mathcal{V}_{\text{HOM,raw}}\frac{1}{(1-\epsilon)^2}\frac{R^2+T^2}{2RT}(1+2\Tilde{g}^{(2)}(0))
\end{equation}

with the classical interferometer visibility $1-\epsilon=0.99$, and the transmission (reflection) $T=0.495$ ($R=0.505$) of the interfering beam splitter. Because of the short lifetime, there is effectively no temporal filtering.

\subsubsection*{Entanglement measurement and density matrix reconstruction}\label{method:entanglement}
The two-photon density matrix in polarisation space is reconstructed by performing cross-correlation measurements between XX and X photons projected on different polarisation states. After filtering, the XX and X photons are projected onto two chosen orthogonal polarisation states using a quarter-wave plate, half-wave plate and a polarising beam splitter (PBS) followed by linear polarisers aligned to the PBS. With this setup, 36 measurements in the $\{\,|H\rangle, |V\rangle, |D\rangle, |A\rangle, |R\rangle, |L\rangle\,\}_X \otimes \{\,|H\rangle, |V\rangle, |D\rangle, |A\rangle, |R\rangle, |L\rangle\,\}_{XX}$ basis are performed. These cross-correlations are evaluated with a time window of \SI{2}{ns}, except for $V=\SI{0.87}{V}$, where \SI{4}{ns} are used because of the longer lifetime.  The density matrix $\ket{\zeta}$ is reconstructed using a maximum likelihood estimation~\cite{james2001}. 
To counteract the random basis change between source and analyser, the density matrix is multiplied with a unitary transformation $U$, where $U$ is found by minimising $||~U\ket{\zeta}-{\ket{\Phi_+}}~||$. This transformation does not influence the degree of entanglement of the state and is solely performed to calculate the fidelity to the $\ket{\Phi_+}$ Bell state $\mathcal{F}_{\ket{\Phi_+}}$.
The uncertainty is calculated by performing seven measurements at the same voltage of \SI{1.1}{V} on the same QD on different days. For each density matrix we extract the concurrence and fidelity and define the resulting standard deviation as the estimated measurement uncertainty. This uncertainty is much larger than the statistical uncertainty of the concurrence and fidelity of a single density matrix, so the latter is neglected.